\documentclass[aps,prl,twocolumn,showpacs,superscriptaddress,10pt]{revtex4-1}
\pdfoutput=1
\usepackage{graphicx}
\usepackage{dcolumn}
\usepackage{bm}
\usepackage{amsmath}
\usepackage{amssymb}
\usepackage{textcomp}
\usepackage{color}

\allowdisplaybreaks[1]

\begin{document}

\title{Picosecond laser filamentation in air}

\author{Andreas~Schmitt-Sody}
\affiliation{Air Force Research Labs, Kirtland Air Force Base, Albuquerque, New Mexico, USA}

\author{Heiko~G.~Kurz}
\affiliation{ Institute of  Quantum Optics, Leibniz University Hannover, Welfengarten 1, 30167 Hannover, Germany}

\author{L. Berg\'e}
\affiliation{CEA-DAM, DIF, F-91297 Arpajon, France}

\author{S. Skupin}
\affiliation{Univ.~Bordeaux - CNRS - CEA, Centre Lasers Intenses et Applications, UMR 5107, 33405 Talence, France}

\author{Pavel~Polynkin}
\email{ppolynkin@optics.arizona.edu}
\affiliation{College of Optical Sciences, The University of Arizona, Tucson, Arizona, USA}

\date{\today}

\begin{abstract}
The propagation of intense picosecond laser pulses in air in the presence of strong nonlinear self-action effects and air ionization is investigated experimentally and numerically. The model used for numerical analysis is based on the nonlinear propagator for the optical field coupled with the rate equations for the production of various ionic species and plasma temperature. Our results show that the phenomenon of plasma-driven intensity clamping, 
which is paramount in femtosecond laser filamentation, holds for picosecond pulses. Furthermore, the temporal pulse distortions are limited and
the pulse fluence is also clamped. The resulting unique feature of the picosecond filamentation regime is the production of a broad, fully ionized air channel, continuous both longitudinally and transversely, which may be instrumental for numerous applications.
\end{abstract}


\maketitle

Ionization of gases is a major application of intense pulsed lasers. 
Two specific interaction regimes have been extensively studied. Ionization by nanosecond (ns) laser pulses (optical breakdown) 
has been investigated since the early 1980's~\cite{Bekefi:PLP:76}. In this context, experimental geometries with tight beam focusing are typically used. 
The physics of the gas response in optical breakdown is very complex~\cite{Raizer:GDP:91,Itikawa:jpcrd:18:23}, 
but the propagation of the ns laser field itself is usually treated in a simple way, e.g.,  
by assuming fixed temporal and spatial distributions for the laser intensity. The second configuration involves intense femtosecond (fs) laser pulses. 
Here, the temporal and spatial dynamics of the optical field are highly nonlinear~\cite{Couairon:pr:441:47,Berge:rpp:70:1633,Champeaux:pre:68:066603,Couairon:jmo:53:75,Odhner:prl:105:125001,Polynkin:pra:87:053829,Kosareva:ol:22:1332}.
However, the complex field dynamics are essentially decoupled from the rich air photochemistry driven by electron-particle collisions, as the latter happen 
on longer time scales than the laser pulse duration.

Compared to the ns and fs cases discussed above, the propagation of either collimated or weakly focused intense picosecond (ps) laser pulses in gaseous media has been studied to a much lesser extent. 
Few works have been published on the subject. All of them utilized ps lasers with rather low pulse energies ($\sim$mJ), operating in the UV~\cite{Schwarz:oc:180:383} or visible~\cite{Mikalauskas:apb:75:899} spectral domains. The generated plasma densities were much lower
than what can be achieved with near-infrared Joule-level ps laser pulses, as we will show below. In the ps regime, both the highly nonlinear propagation of the laser pulse and the complex response of the gas enabled by the electron-molecule and electron-ion collisions occur on comparable timescales. 
This is a very challenging regime to explore, as high energies in the multi-Joule range are needed to drive significant avalanche ionization of the gas.

From the practical angle, studies in this field are motivated by various atmospheric applications 
such as air lasing~\cite{Hemmer:pnas:108:3130}, channeling electrical discharges~\cite{Koopman:jap:44:5328} and microwave beams~\cite{Shen:jap:69:6827}, and lightning control~\cite{Uchida:jot:66:199}. Plasma produced through optical breakdown of air by tightly focused ns laser pulses is dense but spatially localized. By contrast, plasma channels produced through fs laser filamentation in air are extended but short-lived ($<1$\,ns~\cite{Sun:pre:83:046408}) and dilute, with plasma densities of 
$10^{16}$\,--\,$10^{17}$ electrons per cm$^{3}$~\cite{Chen:prl:105:215005}. 
The hybrid fs-ns excitation approach using the so-called igniter-heater scheme
produces both dense and extended plasma channels in air~\cite{Polynkin:apl:99:151103,Polynkin:pra:86:043410}, but thus generated plasma channels are usually fragmented into discrete plasma bubbles. All of the above shortcomings severely limit applications.  

In this Letter, we report results of experiments and numerical simulations on the propagation of intense, weakly focused, ps laser pulses in air. We show that 
all of the above-mentioned drawbacks of the ns, fs and the hybrid fs--ns regimes can be avoided through the use of ps laser pulses, 
promoting robust, continuous and dense plasma columns in air.

Our experiments make use of the Comet laser system which 
is a part of the Jupiter Laser Facility at the Lawrence Livermore National Laboratory in California, USA~\cite{JLF}. Comet is a chirped-pulse amplification chain based on Nd-doped glass. It produces 0.5\,ps transform-limited pulses with maximum energy of 10\,J at the wavelength of 1.053\,{\textmu}m, in a 90\,mm-diameter beam. The laser operates in the single-shot regime (one shot in ten minutes).
Laser pulses can be chirped to up to 20\,ps duration. 
To avoid the degradation of
the 10\,mm-thick fused-silica exit window of the laser's vacuum compressor chamber, the peak laser power must not exceed 1\,TW, which is more than one hundred times 
the critical power for self-focusing in air.
Therefore, for pulses shorter than 10\,ps, the laser energy is reduced accordingly. The laser beam is weakly focused in the ambient air using a 5\,mm-thick, 150\,mm-diameter meniscus lens 
with 3\,m focal length.

\begin{figure}
\includegraphics[width=\columnwidth]{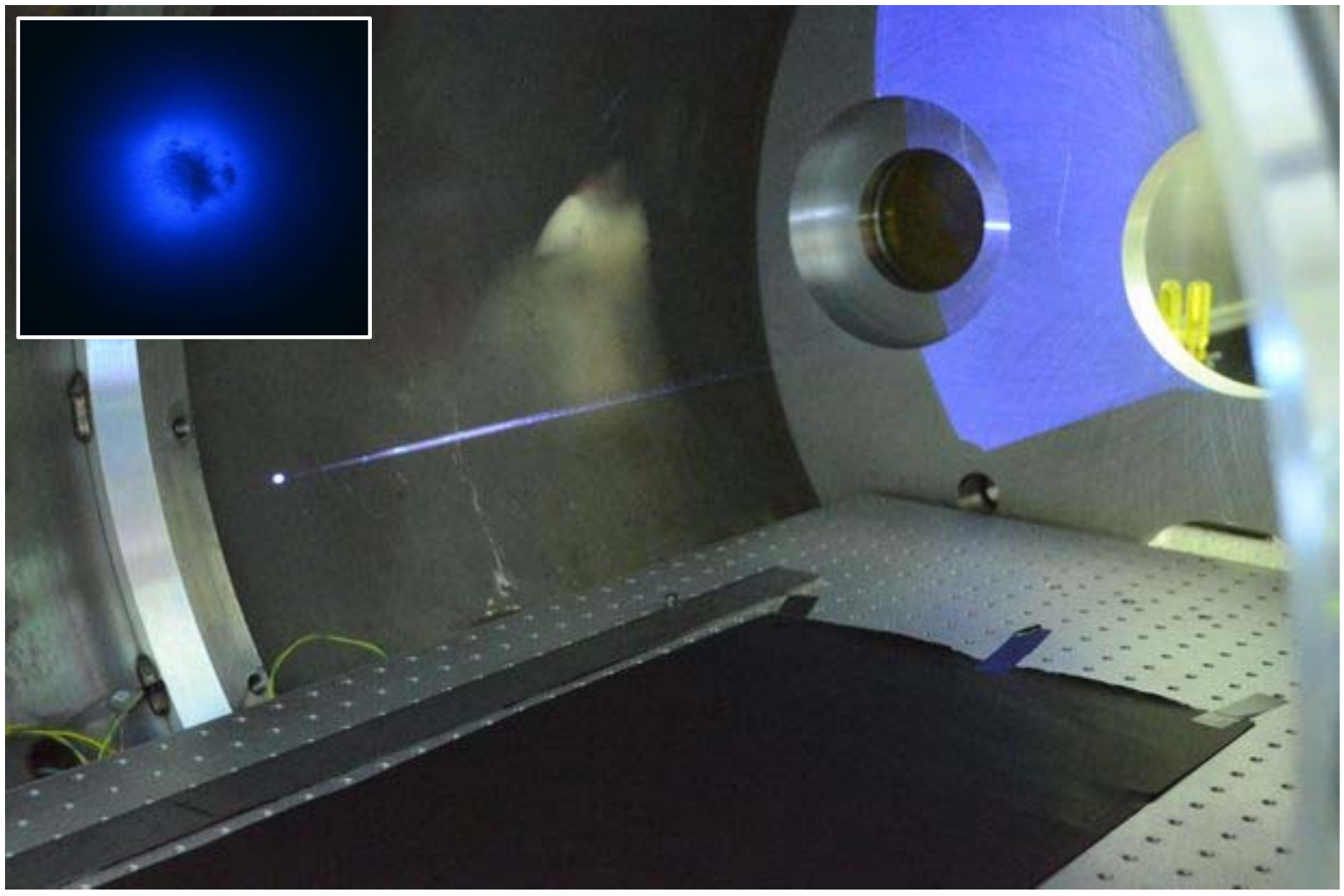}
\caption{
A photograph of the time-integrated fluorescence by the plasma column produced in air by a 10\,J, 10\,ps laser pulse, propagating from right to left and slightly downward.
The inset shows fluorescence induced by the laser beam on a white paper screen placed in the far field. }
\label{photograph}
\end{figure}

In Fig.~\ref{photograph} we show a single-shot photograph of the plasma channel produced by a 10\,J, 10\,ps laser pulse. The $\sim$30\,cm-long plasma string appears to be dense and longitudinally continuous, but the resolution of this image does not guarantee the transverse homogeneity of the plasma.   
The inset 
shows the fluorescence produced by the laser beam in far field, on a white paper screen. The image is taken through a color-glass filter that blocks the overwhelming light at the fundamental laser wavelength of 1.053\,{\textmu}m. Darker areas in the image result from the screening of the trailing edge of the laser pulse by the electron plasma generated on the leading pulse edge. The large dark area in the middle corresponds to the transversely continuous dense plasma on the beam axis. It is surrounded by several peripheral point-like plasma filaments. The production of a large-diameter plasma column that is not fragmented into
multiple filaments may be compared to the "superfilamentation" phenomenon recently discussed in connection with relatively tightly focused fs laser pulses~\cite{Point:prl:112:223902}. 
The difference in the ps case is the important contribution of avalanche ionization, resulting in plasma densities significantly higher than those achieved in fs superfilaments. 

It is well-known that the optical intensity in fs laser filaments in gases is clamped to
a value that is independent of the energy of the laser pulse~\cite{Xu:lp:22:195}. 
The clamping phenomenon is a consequence of the threshold-like dependence of the ionization yield on the optical intensity. Once the ionization threshold is reached, plasma defocusing abruptly stops the further beam collapse. This description is somewhat oversimplified as it does not account for the complex 
temporal pulse dynamics in femtosecond filaments. It is more appropriate to talk about clamping of the time-integrated intensity (i.e. fluence). 
Nevertheless, the intensity clamping concept has been central in the fs filamentation science for quite some time.

\begin{figure}
\includegraphics[width=\columnwidth]{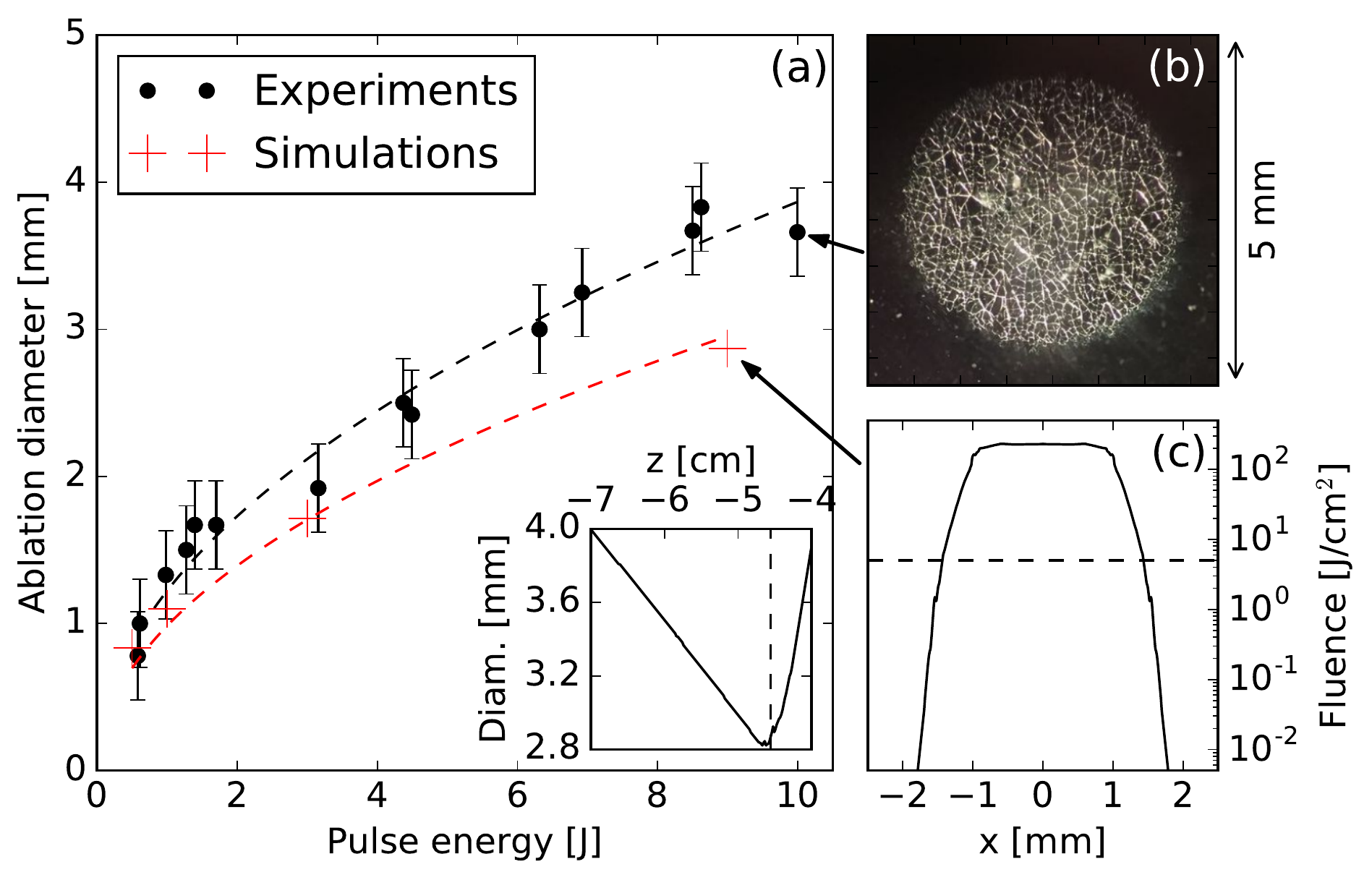}
\caption{
(a) Measured ($\bullet$) and simulated (\textcolor{red}{$+$}) diameters of ablation marks produced by 10\,ps pulses with various energies 
on a glass surface placed at the position of peak air ionization. The dashed curves are fits proportional to the square-root of the pulse energy.
(b) Photograph of the ablation mark produced by a 10\,J, 10\,ps pulse. (c) The simulated fluence profile for a 9\,J, 10\,ps pulse. 
The 5\,J/cm$^{2}$ level corresponding to the ablation threshold fluence for 10\,ps pulse duration is marked with a dashed line. 
The inset in (a) shows the simulated beam diameter at the 5\,J/cm$^{2}$ level, vs. the distance from the geometrical focus, for a 9\,J, 10\,ps pulse.}
\label{burndiam}
\end{figure}

In order to evaluate the fluence of the ps laser field in the interaction zone we produce the single-shot ablation of a front surface of a glass slide placed along the beam path close to the position of maximum plasma production. This position is determined as the point of the brightest plasma fluorescence on a photographic 
image of the plasma string (similar to Fig.~\ref{photograph}).
The ablation pattern marks a circular region with a sharp boundary, inside which the laser fluence exceeds the value of the ablation threshold 
for the glass material (see Fig.~\ref{burndiam}). We have separately measured the ablation threshold fluence for the particular type of glass we used (borofloat microscope slides), 
for near-infrared laser pulses with durations in the range from 1 to 10\,ps. 
Our results are consistent with the results for fused silica glass reported in~\cite{Stuart:josab:13:459} and show a steady growth of the ablation threshold fluence 
with pulse duration in that range, from $2.5$ to 5\,J/cm$^{2}$.

\begin{figure}
\includegraphics[width=\columnwidth]{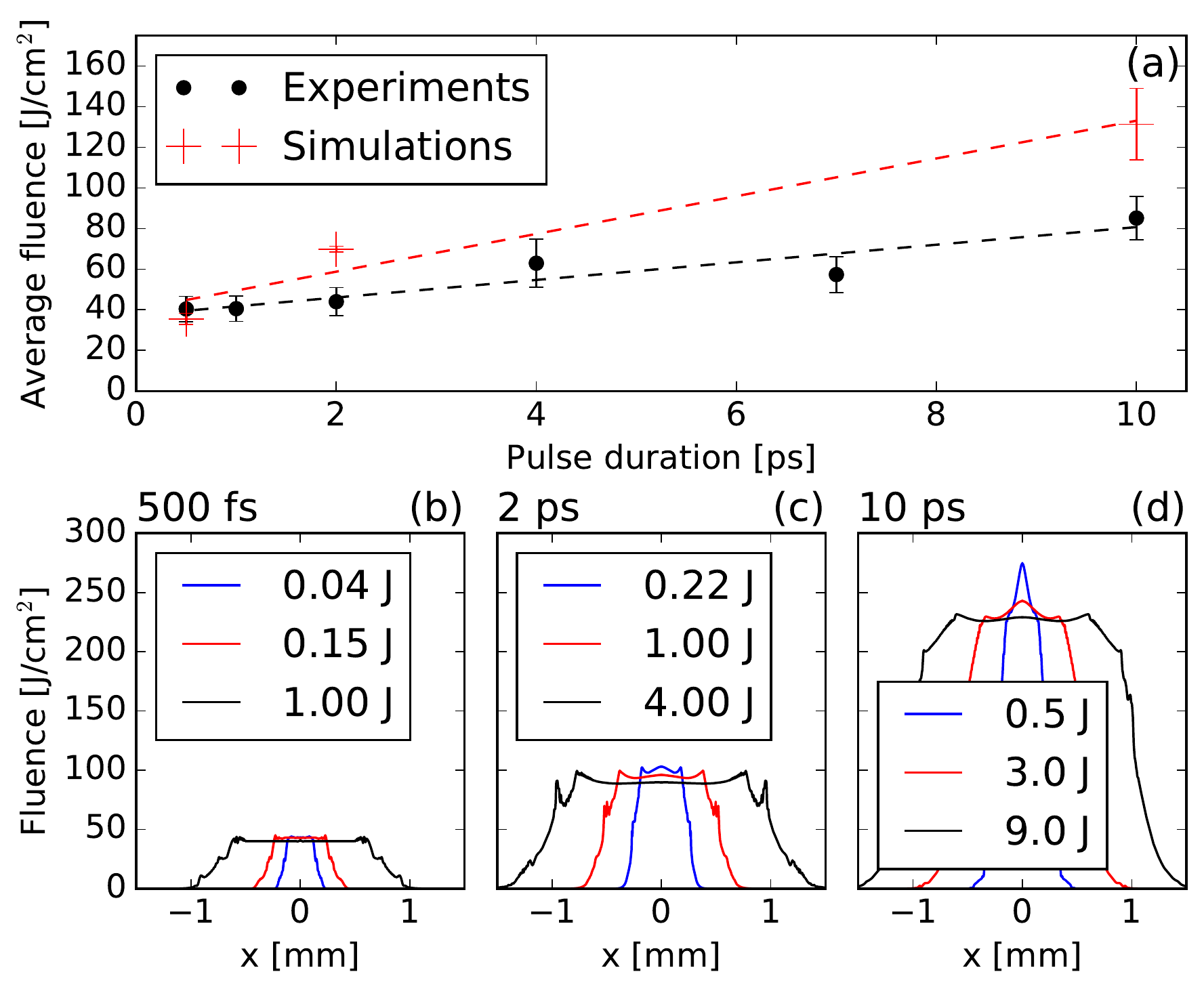}
\caption{
(a) Measured ($\bullet$) and simulated (\textcolor{red}{$+$}) average laser fluence at the position of peak air ionization, vs. pulse duration. 
At any given pulse duration, the average fluence is clamped, i.e., approximately independent of the pulse energy. 
(b)-(d) Computed fluence profiles at the positions of peak ionization, for various pulse configurations, showing fluence clamping.}
\label{clamping}
\end{figure}

Data for the diameter of the ablation mark, for a 10\,ps pulse, is shown in Fig.~\ref{burndiam}. 
The corresponding area of the mark scales linearly with the pulse energy, as evidenced by the agreement with the dashed regression curve.
Thus the average fluence computed as the ratio of the pulse energy to the ablation area is approximately constant or clamped with respect to the pulse energy.   
We have conducted these measurements using laser pulses with various durations and found that the average laser fluence was clamped in all cases. 
The actual value of the clamped fluence was found to be a growing function of the pulse duration, as shown in Fig.~\ref{clamping}(a).  

For numerical simulations of our experiments we have devised a model that combines a unidirectional nonlinear pulse propagator for the laser field amplitude $U$~\cite{Kolesik:prl:89:283902} with a material response module.
The unidirectional treatment is valid as the estimated peak reflectivity of the plasma is less than $10^{-4}$, even assuming a complete single ionization of all air molecules.
Our model accounts for multi-photon and impact single ionization of oxygen and nitrogen molecules
with the initial densities of $\rho_{N_2}^{(0)} = 2.2 \times 10^{19}$\,cm$^{-3}$ and $\rho_{O_2}^{(0)} = 5.4 \times 10^{18}$\,cm$^{-3}$, respectively. 
Direct photoionisation via Perelomov, Popov and Terent'ev (PPT) theory~\cite{Perelomov:spjetp:23:924} is included for oxygen molecules only, as the ionization potential of oxygen is lower than that of nitrogen. 
Collisional ionization is included for both neutral species. To that end, we compute the inverse bremsstrahlung heating of free electrons using the electron temperature-dependent collision rates $\nu_{X}(T_e)$ in the respective collision cross-sections $\sigma_X = e^2 \nu_{X}(T_e)/m_e \epsilon_0 n_0 c \omega_0^2$~\cite{Huba:NRL:13} (subscript $X$ indexes different molecular and ionic species: $N_2$, $N_2^+$, $O_2$ and $O_2^+$).
Our nonlinear propagator model reads as follows:
\begin{align}
\begin{split}
\frac{\partial}{\partial z} U &  =  \frac{i}{2k_0} \hat{T}^{-1}\nabla_{\perp}^2 U + i \hat{\mathcal{D}} U
 + i \frac{\omega_0}{c} n_2\hat{T} R*|U|^2 U \\
& \quad  -  i \frac{k_0}{2n_0^2 \rho_c} \hat{T}^{-1} \rho_e U - \frac{\sigma_e}{2} \rho_e U \\
& \quad  -  \frac{ E_{O_2} W_{O_2}^{\rm{PPT}} (|U|^{2}) }{2|U|^{2}}(\rho_{O_2}^{(0)} - \rho_{O_2^+}) U ,  \label{eq:UPPE}
\end{split} \\
R(t) &  = \frac{1}{2}\delta(t) + \frac{1}{2}\Theta(t)\frac{\tau_1^2+\tau_2^2}{\tau_1\tau_2^2}e^{-t/\tau_2}\sin(t/\tau_1), \label{eq:raman} \\
\frac{\partial}{\partial t} \rho_{O_2^+} & = W_{O_2}^{\rm{PPT}}(|U|^{2}) (\rho_{O_2}^{(0)} - \rho_{O_2^+})
+ \frac{\sigma_{O_2}}{E_{O_2}} \rho_e |U|^2, \\
\frac{\partial}{\partial t} \rho_{N_2^+} & = \frac{\sigma_{N_2}}{E_{N_2}} \rho_e |U|^2, \label{eq:n2}\\
\begin{split}
\frac{\partial}{\partial t} T_e & = \frac{2}{3k_B}\sigma_{e}|U|^{2} -\left( \frac{T_e}{E_{O_2}} + \frac{2}{3 k_B} \right) \sigma _{O_2} \vert U \vert ^{2} \\
& \quad 
-\left( \frac{T_e}{E_{N_2}} + \frac{2}{3 k_B} \right) \sigma _{N_2} \vert U \vert ^{2}.
\end{split}
\end{align}
Here, we use the conventional notation for fundamental constants;
$\omega_0$ is the laser angular frequency, $n_0$ is the corresponding linear refractive index of air, $k_0=\omega_0n_0/c$ is the wavenumber, $\rho_c$ is the critical plasma density and $\rho_e = \rho_{O_2^+} + \rho_{N_2^+}$ is the free-electron density. The operator $\hat{T}$ accounts for the self-steepening and space-time focusing effects~\cite{Brabec:prl:78:3282}. Linear dispersion of air is included via operator $\hat{\mathcal{D}}$~\cite{Peck:josa:62:958}.
The nonlinear refractive index of air $n_2 \, = \, 2\times 10^{-19}$\,cm$^2$/W is equally partitioned into the instantaneous and delayed contributions according to 
Eq.~(\ref{eq:raman}) with $\tau_1 =  60$\,fs and $\tau_2  =  80$\,fs~\cite{Pitts:josab:21:2008}. 
$E_{N_2}$ and $E_{O_2}$ are the ionization potentials of the neutral nitrogen and oxygen molecules, respectively.
The compound collision cross-section is defined as $\sigma_{e}=\sigma_{O_2^+}+\sigma_{N_2^+} + \sigma_{O_2}+\sigma_{N_2}$.
Electron recombination and attachment to neutral molecules are neglected as the time scales for those processes are much longer than the pulse duration. 
Note that accounting for the depletion of neutrals in the Kerr term and photoionization of $N_2$ using the corresponding PPT rate in Eq.~(\ref{eq:n2}) do not appreciably change the propagation dynamics.

Results of our numerical simulations are shown in Figs.~\ref{burndiam} and \ref{clamping}, together with their experimental counterparts. The simulations used here utilize an axially-symmetric implementation of the model discussed above. 
In order to mimic the experimental conditions we use Gaussian temporal $\propto\exp(-t^2/t_0^2)$ and super-Gaussian spatial $\propto\exp(-r^4/w_0^4)$ 
profiles for the input laser field.
Simulations start 8\,cm before the position of the geometrical focus, where intensities 
do not exceed 10\,TW/cm$^2$ and ionization is still negligible. 
The code propagates the focused pulse, finds the longitudinal position of maximum air ionization and, at that position, computes the diameter of the area inside which the fluence exceeds the ablation threshold value 
(e.g., 5\,J/cm$^{2}$ for the case of 10\,ps pulse). Simulation results for the size of ablation agree with experiments within about 25\%.
The experimental ablation marks may be larger than the computed ones because of (i) the beam defects that are not accounted for in the model; (ii) the uncertainty of the longitudinal placement of the glass sample, estimated as $\pm 0.5$\,cm. 
From the simulated beam diameter at the 5\,J/cm$^{2}$ level vs. longitudinal position shown in the inset in Fig.~\ref{burndiam}(a), it is clear that the placement uncertainty may affect the measurement significantly. In contrast, the uncertainty of the value of the ablation threshold should not significantly affect the computed diameter of the ablation mark due to the sharp spatial gradients of the fluence profile. 
In Fig.~\ref{clamping}(a) we show the average fluence, defined here as a ratio of the pulse energy to the area with fluence 
above the corresponding ablation threshold value. Numerical and experimental data agree reasonably well and both show a steady growth of the average fluence 
with pulse duration. The inspection of the computed fluence profiles [Fig.~\ref{clamping}(b-d)] confirms fluence clamping.

\begin{figure*}
\includegraphics[width=\textwidth]{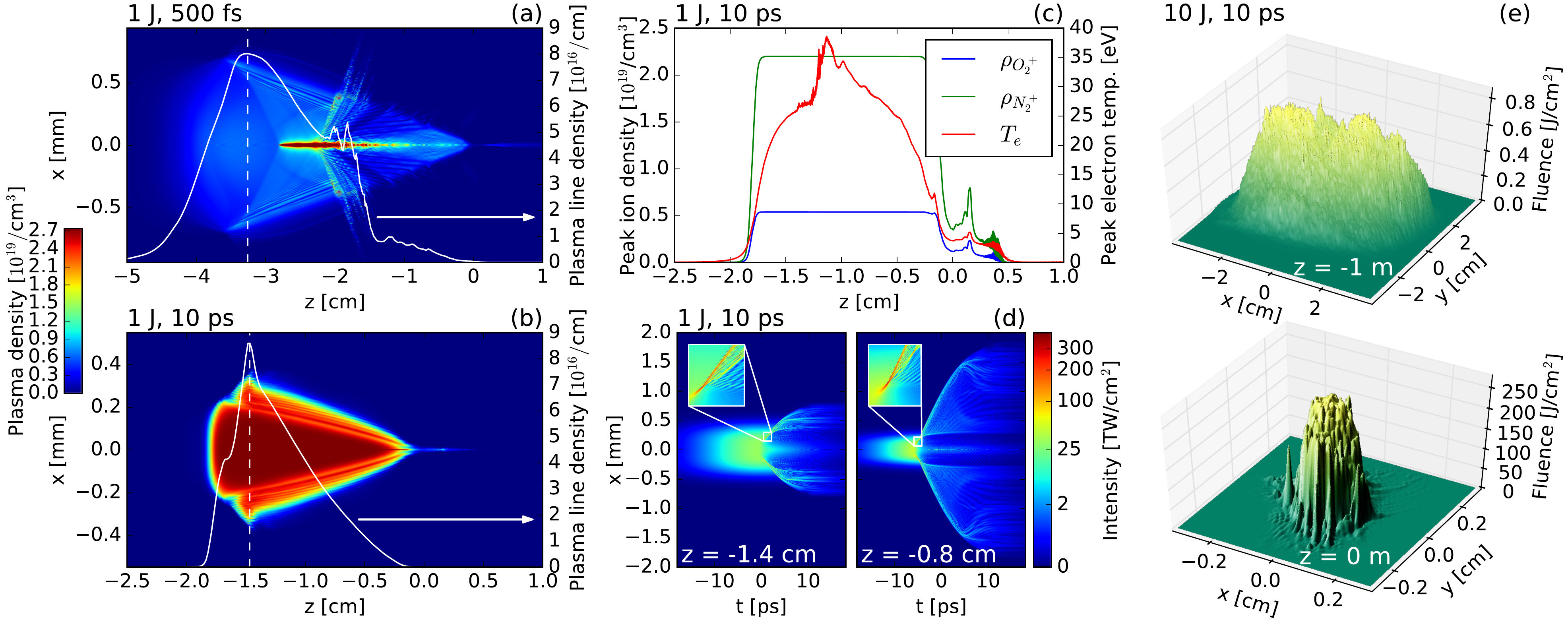}
\caption{(a) Simulated plasma channel produced by a 500\,fs, 1\,J laser pulse. The longitudinal position of the peak plasma production (the maximum of the linear plasma density $\int \rho_e(r,z,t) rdr$ for $t$ larger than the pulse duration) is marked by the dashed white line. (b) Same for a 10\,ps pulse with same energy. (c) Peak ionic densities and electron temperature vs. propagation distance for the same pulse as in (b). (d) Snapshots of the corresponding spatio-temporal intensity profiles. 
(e) Input and output fluence profiles computed using the time-integrated propagation model and a 10\,J, 10\,ps noisy input beam (see text for details). 
All longitudinal coordinates are shown with respect to the linear geometrical focus of the focusing lens.}
\label{simulations}
\end{figure*}

Simulations show a complete single ionization of both oxygen and nitrogen molecules in the ps filament. Plasma density exceeds $10^{19}$\,cm$^{-3}$ for a wide range of pulse energies and durations. Figures~\ref{simulations}(a,b) show the distributions of plasma produced by 500\,fs and 10\,ps pulses, each with 1\,J of energy. While the shorter pulse produces plasma near the beam axis only, the 10\,ps pulse produces a much larger, homogeneous plasma channel with complete single ionization of air molecules. Electron temperatures up to 35\,eV are achieved 
with 1\,J pulses in the region of complete ionization, as shown in Figure~\ref{simulations}(c). With 10\,J pulses, electron temperatures as high as 100\,eV are observed.
The inspection of the spatio-temporal intensity profiles [two snapshots are shown in Fig.~\ref{simulations}(d)] reveals that a complete single ionization of air, predominantly through collisional ionization, occurs on the leading edge of the pulse, which remains relatively intact with intensity clamped at the level below 100\,TW/cm$^2$. 

Simulations discussed so far utilized an axially-symmetric implementation of our model. Thus based on this data we cannot rule out the 
possible onset of multiple filamentation, which may break the plasma channel in the transverse plane. To investigate the transverse uniformity of the plasma column, we conduct another set of simulations employing a time-integrated version of Eq.~(\ref{eq:UPPE}). 
Here we use a noisy input fluence profile whose small-scale fluctuations efficiently seed the transverse modulation instability of the beam~\cite{Berge:prl:92:225002,Skupin:pre:70:046602}. 
Figure~\ref{simulations}(e) shows the fluence profile produced by such a simulation for a 10\,J, 10\,ps pulse. The simulation is started 1\,m before the position of the geometrical focus, where intensities are below 70\,GW/cm$^2$ and nonlinear effects are negligible. The computed fluence profile at the focus shows no sign of multiple filamentation at the beam center, confirming analytical estimates that suggest that the growth rates of the transverse modulation instability~\cite{Bespalov:jetp:3:307} are too small for the instability to be effective on the length scale of our experiment. We see, however, several peripheral smaller plasma filaments, compatible with the experimental observations (see inset in Fig.~\ref{photograph}).

High plasma densities predicted by simulations are experimentally verified by the measurement using a capacitive plasma probe 
\cite{Meyerand:prl:11:401,Polynkin:apl:101:164102}. The probe that we use has two 5\,cm--long electrodes separated by a distance of 3\,cm 
and charged to a voltage of 50\,V. The signal produced by the probe in our experiments is five to six orders of magnitude higher than the signal from the same probe applied to a regular femtosecond plasma filament with an independently known plasma density between $10 ^{16}$ and $10 ^{17}$\,cm$^{-3}$ \cite{Chen:prl:105:215005}. Accounting for the difference 
between the total plasma volumes in the two cases, these measurements confirm a complete or nearly complete ionization of air in the interaction zone.

In conclusion, we have investigated the propagation of Terawatt ps laser pulses in air by means of experiments and numerical simulations. 
Our model is based on a unidirectional pulse propagator coupled to a material response module that accounts for multi-photon and impact ionization of air molecules, as well as heating of electron plasma. We report good agreement between experimental and numerical results. Our major finding is the possibility to produce a broad, fully ionized 
and longitudinally and transversely continuous air channel, that can be instrumental for applications. 

This work was supported by the US Defense Threat Reduction Agency under program HDTRA 1-14-1-0009 and by the US Air Force Office of Scientific Research under program FA9550-12-1-0482. 
The use of the Jupiter Laser Facility was supported by the US Department of Energy, Lawrence Livermore National Laboratory, under Contract No.\ DE-AC52-07NA27344. 
Numerical simulations were performed at M\'eso\-centre de Calcul Intensif Aquitain (MCIA), Grand
Equipement National pour le Calcul Intensif (GENCI, grant
no.~2015-056129), and Partnership for Advanced Computing in Europe
(PRACE, grant no.~2014112576). LB thanks Patrick Combis for discussions on ablation thresholds in glasses.


%

\end{document}